\documentclass[aip,amsmath,amssymb,floatfix,citeautoscript,reprint]{revtex4-1}
\usepackage{cancel}
\usepackage{amsmath}
\usepackage{graphicx}
\usepackage{bm}
\usepackage{physics}
\usepackage[version=3]{mhchem}
\usepackage{txfonts}

\usepackage{color}
\usepackage[usenames,dvipsnames]{xcolor}
\definecolor{myblue}{rgb}{0,0,1}
\usepackage[breaklinks=true,colorlinks=true,linkcolor=myblue,urlcolor=myblue,citecolor=myblue]{hyperref}

\begin{document}
\title{Improved Fast Randomized Iteration Approach to Full Configuration Interaction}

\author{Samuel M. Greene}
\affiliation{Department of Chemistry, Columbia University, New York, New York 10027, United States}
\author{Robert J. Webber}
\author{Jonathan Weare}
\email{weare@cims.nyu.edu}
\affiliation{Courant Institute of Mathematical Sciences, New York University, New York, New York 10012, United States}
\author{Timothy C. Berkelbach}
\email{tim.berkelbach@gmail.com}
\affiliation{Department of Chemistry, Columbia University, New York, New York 10027, United States}
\affiliation{Center for Computational Quantum Physics, Flatiron Institute, New York, New York 10010, United States}

\begin{abstract}
We present three modifications to our recently introduced fast randomized iteration method for full configuration interaction (FCI-FRI) and investigate their effects on the method's performance for Ne, \ce{H2O}, and \ce{N2}. The initiator approximation, originally developed for full configuration interaction quantum Monte Carlo, significantly reduces statistical error in FCI-FRI when few samples are used in compression operations, enabling its application to larger chemical systems. The semi-stochastic extension, which involves exactly preserving a fixed subset of elements in each compression, improves statistical efficiency in some cases but reduces it in others. We also developed a new approach to sampling excitations that yields consistent improvements in statistical efficiency and reductions in computational cost. We discuss possible strategies based on our findings for improving the performance of stochastic quantum chemistry methods more generally.
\end{abstract}

\maketitle

\section{Introduction}
Strong correlation among electrons in many materials gives rise to unique properties that are potentially of high value in applications, e.g. high magnetic susceptibility, superconductivity, or catalytic behavior~\cite{Rao1989, Edelstein2003, Assaad2006, Krcha2014, Amusia2015, Zheng2017, Vogiatzis2019}. These materials are often not well understood from a theoretical standpoint due to the high cost of numerically solving the Schr\"odinger equation for their constituent electrons~\cite{Zhang2004, Booth2013, Tubman2016, Schwarz2017}. A number of methods are therefore being developed to accurately approximate its solution at an affordable cost~\cite{Anisimov1991, Liechtenstein1995, Mazziotti2011, Lyakh2011, Zhang2016, Hofstetter2018}. We recently introduced a class of stochastic methods, termed FCI-FRI~\cite{Greene2019}, for approximating the ground-state eigenvector of the electronic Hamiltonian matrix expressed in a discrete basis of Slater determinants, i.e. the full configuration interaction (FCI) matrix~\cite{Knowles1984}. 

Deterministic (i.e. non-stochastic) iterative linear algebra methods, e.g. the Lanczos~\cite{Lanczos1950} or Jacobi-Davidson~\cite{Davidson1975, Sleijpen1996} algorithms, are conventionally used to calculate low-energy eigenvalues and eigenvectors of the FCI matrix. These methods involve calculating a series of matrix-vector products. Because the dimension of the FCI matrix increases combinatorially with the number of electrons and single-particle basis size, the cost of these calculations can be prohibitively expensive even for relatively small chemical systems. A variety of methods, including FCI-FRI, address this challenge by zeroing matrix and vector elements~\cite{Booth2009, Booth2010, Booth2014, Blunt2014, Shepherd2014, Blunt2015, Blunt2015a, Alavi2016, Holmes2016, Lu2017, Sharma2017, Wang2019}. When implemented using sparse linear algebra tools, this approach can enable significant gains in computational efficiency over those that do not leverage sparsity.


Both FCI-FRI and the more general Fast Randomized Iteration (FRI) framework on which it is based~\cite{Lim2017} were motivated in large part by the variety of quantum Monte Carlo (QMC) methods developed over the past several decades~\cite{Barker1979, Hammond1994, CalandraBuonaura1998, Foulkes2001, Maksym2005, Needs2010, Austin2012, Scott2017, Motta2018} and in particular by the FCIQMC methods developed over the past 11 years~\cite{Booth2009, Booth2010, Booth2014, Blunt2014, Shepherd2014, Blunt2015, Blunt2015a, Alavi2016}. FCIQMC can be understood as an implementation of the iterative power method in which matrix-vector multiplication operations are simulated via the dynamics of ``walkers'' that transition among randomly selected Slater determinant basis states, with probabilities that depend on matrix elements~\cite{Vigor2016}. 
FCI-FRI generalizes this viewpoint of individual interacting walkers 
by representing the solution as a single vector that evolves according to the usual power method 
with randomly introduced sparsity to reduce the cost of matrix-vector multiplication.  The resulting methods have many fundamental similarities to FCIQMC. Nonetheless, this change in perspective has implications for algorithm design.  Among these is the possibility of introducing correlations to selections that are performed independently in FCIQMC, as well as increased control over the degree of sparsity enforced at various stages of the algorithm. The additional correlations reduce the statistical error in each iteration, as measured by the discrepancy between the updated FCI-FRI vector and the corresponding deterministic matrix-vector product. This leads
to significant reductions in overall statistical error, as demonstrated by our previous results for several small chemical systems~\cite{Greene2019}.  
The accuracy of any FCI-FRI calculation can be systematically improved by retaining more nonzero elements in each iteration, generally at increased computational cost.  A central focus of our ongoing work is reducing the computational cost and corresponding statistical error of these methods in order to enable their application to larger systems of interest in chemistry and physics.

Since the development of the original FCIQMC method, a number of modifications have been introduced to improve its performance. 
For example, the initiator approximation reduces the large statistical error observed when few walkers are used, thereby enabling the application of FCIQMC methods to significantly larger chemical systems~\cite{Cleland2010, Cleland2012}. This approximation involves zeroing Hamiltonian matrix elements in each iteration on the basis of their signs relative to those of elements in the vector being multiplied. A later extension involves calculating a perturbative correction to the energy from these zeroed elements~\cite{Blunt2018}. The semi-stochastic adaptation allows for the exact preservation of a predefined set of matrix and vector elements, which reduces the degree of randomness introduced in each iteration~\cite{Petruzielo2012, Blunt2015b}. 
Using improved ``excitation generators,'' i.e. approaches to selecting the probabilities governing transitions among Slater determinants, enables reductions in statistical error~\cite{Holmes2016, Neufeld2019}. These extensions are mostly independent of each other and therefore can be combined for compounded improvements in accuracy and performance.

This article serves two purposes. First, it demonstrates that these modifications originally introduced in an FCIQMC context are applicable to FCI-FRI methods more generally. By demonstrating their compatibility, we suggest the possibility of new methods that combine the best features of existing FCIQMC and FCI-FRI methods for improved computational performance and reduced statistical error. We focus in particular on the initiator and semi-stochastic modifications from FCIQMC as examples of modifications compatible with FCI-FRI. Second, we illustrate the value of working within the FRI framework by introducing an improved Hamiltonian matrix factorization (analogous to an excitation generator), the development of which is facilitated by the framework's generality. The effects of each of these three modifications on  performance and accuracy are evaluated through numerical tests on small chemical systems. 

Although this analysis is applicable to any of the FCI-FRI methods described in our previous work~\cite{Greene2019}, we focus in particular on the best-performing method, namely ``systematic FCI-FRI.'' We thus provide a summary of this method in Section \ref{sec:sysFRI}. Section \ref{sec:initiator} then casts the initiator approximation into the FRI framework.
Like FCIQMC, systematic FCI-FRI exhibits poor convergence behavior when too few nonzero elements are retained in the vector in each iteration (analogous to using few walkers in FCIQMC). We find that the initiator approximation improves the convergence of FCI-FRI in this regime. 
In Section \ref{sec:semiStoch}, we discuss the potential benefits of a semi-stochastic implementation of FCI-FRI but find that it does not consistently improve performance for all systems tested.
 Section \ref{sec:newHB} describes our alternative Hamiltonian matrix factorization
suited for use in FCI-FRI. 
In Section \ref{sec:frivsFCIQMC}, we compare the performance of systematic FCI-FRI and FCIQMC when the initiator approximation is applied to both. Without our new Hamiltonian factorization, initiator FCI-FRI is 2.4 to 15 times more statistically efficient than initiator FCIQMC, and with it, it is up to 29 times more statistically efficient. Finally, in Section \ref{sec:concl}, we summarize our main conclusions and present some preliminary results for more challenging chemical systems, namely one in a larger basis and one with stronger correlation.

\section{The Systematic FCI-FRI Method for Approximating the Ground-State Eigenvalue}
\label{sec:sysFRI}
The systematic FCI-FRI method is a stochastic implementation of the power method, applied to approximate the ground-state eigenvector of the FCI matrix $\mathbf{H}$, expressed in a basis of Slater determinants with $N$ electrons in $M$ orbitals~\cite{Greene2019}. The random vector calculated at each iteration, termed an \textit{iterate}, is denoted $\mathbf{v}^{(\tau)}$, with $\tau$ indicating the iteration index. Each iteration involves applying a sequence of operations to generate the next iterate $\mathbf{v}^{(\tau + 1)}$ by approximating the matrix-vector product $\mathbf{P}^{(\tau)} \mathbf{v}^{(\tau)}$, where
\begin{equation}
\mathbf{P}^{(\tau)} = \mathbf{1} - \varepsilon \left( \mathbf{H} - S^{(\tau)} \mathbf{1} \right)
\end{equation}
and $S^{(\tau)}$ is chosen to approximate the ground-state energy. The deterministic power method is discussed in more detail in Appendix \ref{sec:powMeth}. This section describes how stochastic compression (i.e. randomly zeroing vector elements) can be applied to reduce the cost of matrix-vector multiplication in each iteration, and how the ground-state energy and its associated statistical error can be estimated from the resulting random iterates.



\subsection{Stochastic Vector Compression}
\label{sec:vecComp}
Introducing zeros into vectors
facilitates the use of sparse linear algebra tools, in which only nonzero elements are stored and manipulated in computer memory. 
Although there are a variety of approaches to stochastic compression~\cite{Lim2017}, we focus here on the specific approach
used in systematic FCI-FRI. When applied to a generic vector $\mathbf{x}$, this scheme ensures that the resulting vector, $\Phi(\mathbf{x})$, has at most $m$ elements, where $m$ is a user-specified parameter. 
Although each element of $\Phi (\mathbf{x})$ is (potentially) random, its expectation value is equal to the corresponding element in $\mathbf{x}$:
\begin{equation}
\label{eq:compDef}
\text{E} \left[ \Phi(\mathbf{x})_i \right] = x_i
\end{equation}

The first step in this scheme involves constructing a subspace $\mathcal{D}$, within which elements of $\mathbf{x}$ are preserved exactly, i.e.
\begin{equation}
\Phi(\mathbf{x})_i = x_i, i \in \mathcal{D}
\end{equation}
$\mathcal{D}$ consists of the $\rho$ largest-magnitude elements of $\mathbf{x}$. If $\mathbf{s}$ is the vector that sorts the elements of $\mathbf{x}$ in order of decreasing magnitude, i.e. $|x_{s_j}| \geq |x_{s_{j+1}}|$, then $\rho$ is the minimum integer value of $h$ for which
\begin{equation}
\label{eq:rhoCriterion}
(m - h) |x_{s_{h+1}}| \leq \sum_{j=h+1}^{||\mathbf{x}||_0} |x_{s_j}|
\end{equation}
where $||\mathbf{x}||_0$ denotes the number of nonzero elements in $\mathbf{x}$. If $m \geq ||\mathbf{x}||_0$, this criterion naturally specifies that all nonzero elements of $\mathbf{x}$ are in $\mathcal{D}$. The largest-magnitude elements of $\mathbf{x}$ can be selected one by one, each in $\mathcal{O}(\log ||\mathbf{x}||_0)$ time, by first constructing a binary heap in $\mathcal{O}(||\mathbf{x}||_0)$ time. This avoids the need to explicitly sort elements of $\mathbf{x}$ by magnitude. 

Elements not in $\mathcal{D}$ are in the subspace denoted as $\mathcal{S}$.  The second step in this compression scheme involves randomly selecting the $(m - \rho)$ elements in $\mathcal{S}$ that will be nonzero in the compressed vector and zeroing the remaining elements. Details of this procedure, including an explanation of how correlations among sampled elements are enforced, can be found in ref \citenum{Greene2019}. This particular combination of exact preservation and correlated sampling
provably minimizes the statistical error in $\Phi(\mathbf{x})$, measured as E$\left[ || \Phi(\mathbf{x}) - \mathbf{x} ||_2^2 \right]$, subject to the constraint of $m$ nonzero elements in $\Phi(\mathbf{x})$~\cite{Webber2020}.


\subsection{Hamiltonian Matrix Factorizations}
\label{sec:Hfac}

The simplest application of this compression scheme to the power method involves compressing each iterate and then multiplying the resulting vector by the matrix $\mathbf{P}^{(\tau)}$ to obtain the next iterate, i.e.
\begin{equation}
\mathbf{v}^{(\tau + 1)} = \mathbf{P}^{(\tau)} \Phi\left(\mathbf{v}^{(\tau)} \right)
\end{equation}
The computational cost of this calculation is dominated by matrix-vector multiplication. The matrix $\mathbf{P}^{(\tau)}$ has the same dimensions and sparsity structure as $\mathbf{H}$, so each of its columns has $\mathcal{O}(N^2 V^2)$ nonzero elements, where $V = M-N$. The cost of performing this matrix-vector multiplication using an efficient sparse linear algebra scheme is therefore $\mathcal{O}(N^2 V^2 m)$, where $m$ is the number of nonzero elements in the compressed vector. Although this is significantly more favorable than a scheme that does not use stochastic compression, the value of $m$ required for accurate results precludes application to many systems of interest in chemistry. Due to this challenge, we do not consider this method further in this paper and focus on methods with reduced cost, described below. However, results obtained by applying this method to the Ne atom in the aug-cc-pVDZ basis, a system with 8 electrons in 22 orbitals, were presented in ref \citenum{Greene2019}.

In order to reduce this cost, we employ a scheme in which iterates are not multiplied by $\mathbf{P}^{(\tau)}$ directly. Instead,
$\mathbf{P}^{(\tau)}$ is partitioned into a sum of two matrices:
\begin{equation}
\mathbf{P}^{(\tau)} = \mathbf{P}^{(\tau)}_\text{diag} + \mathbf{P}_\text{off-diag}
\end{equation}
where $\mathbf{P}^{(\tau)}_\text{diag}$ and $\mathbf{P}_\text{off-diag}$ contain the diagonal and off-diagonal elements of $\mathbf{P}^{(\tau)}$, respectively. This partitioning is motivated by the fact that diagonal elements of $\mathbf{P}^{(\tau)}$ generally have greater magnitudes than those of off-diagonal elements. Only $\mathbf{P}^{(\tau)}_\text{diag}$ varies in each iteration due to the dependence of its elements on the energy shift (eq \ref{eq:Pdef}); elements of $\mathbf{P}_\text{off-diag}$ are constant. $\mathbf{P}_\text{off-diag}$ is exactly factored into a product of six matrices, each of which has $\mathcal{O}(N)$, $\mathcal{O}(V)$, or $\mathcal{O}(1)$ elements per column. The matrix-vector product $\mathbf{P}_\text{off-diag} \Phi(\mathbf{v}^{(\tau)})$ is approximated by multiplying $\Phi(\mathbf{v}^{(\tau)})$ by each of these six matrices in sequence. Before each multiplication operation, the vector is compressed to $m$ nonzero elements, which limits the CPU cost and memory requirements of performing multiplication to $\mathcal{O}(Nm)$, $\mathcal{O}(Vm)$, or $\mathcal{O}(m)$, depending on the number of nonzero elements in the columns of the matrix. The product $\mathbf{P}^{(\tau)}_\text{diag} \Phi(\mathbf{v}^{(\tau)})$ is calculated directly at $\mathcal{O}(m)$ cost and added to the vector approximating $\mathbf{P}_\text{off-diag} \Phi(\mathbf{v}^{(\tau)})$ to obtain the next iterate. This procedure is summarized in Table \ref{tab:FRIsteps}.

\begin{table}
\begin{tabular}{l l}
1. Vector compression: & $\Phi(\mathbf{v}^{(\tau)})$ \\
2. Matrix-vector multiplication: & $\mathbf{q}^{(\tau, 1)} = \mathbf{Q}^{(1)} \Phi(\mathbf{v}^{(\tau)})$ \\
3. Vector compression: & $\Phi(\mathbf{q}^{(\tau, 1)})$ \\
4. Matrix-vector multiplication: & $\mathbf{q}^{(\tau, 2)} = \mathbf{Q}^{(2)} \Phi(\mathbf{q}^{(\tau, 1)})$ \\
5. Vector compression: & $\Phi(\mathbf{q}^{(\tau, 2)})$ \\
$\vdots$ \\
10. Matrix-vector multiplication: & $\mathbf{q}^{(\tau, 5)} = \mathbf{Q}^{(5)} \Phi(\mathbf{q}^{(\tau, 4)})$ \\
11. Vector compression: & $\Phi(\mathbf{q}^{(\tau, 5)})$ \\
12. Matrix-vector multiplication: & $\mathbf{q}^{(\tau, 6)} = \mathbf{B} \Phi(\mathbf{q}^{(\tau, 5)})$ \\
13. Matrix-vector multiplication: & $\mathbf{d}^{(\tau)} = \mathbf{P}^{(\tau)}_\text{diag} \Phi(\mathbf{v}^{(\tau)})$ \\
14. Vector addition: & $\mathbf{v}^{(\tau + 1)} = \mathbf{q}^{(\tau, 6)} + \mathbf{d}^{(\tau)}$
\end{tabular}
\caption{The sequence of steps involved in each iteration of the systematic FCI-FRI method. This formulation is based on the equivalence between the matrices $(\mathbf{P}^{(\tau)}_\text{diag} + \mathbf{B} \mathbf{Q}^{(5)} \mathbf{Q}^{(4)} \mathbf{Q}^{(3)} \mathbf{Q}^{(2)} \mathbf{Q}^{(1)})$ and $\mathbf{P}^{(\tau)}$.}
\label{tab:FRIsteps}
\end{table}


There are several different approaches to factoring $\mathbf{P}_\text{off-diag}$~\cite{Booth2014, Holmes2016, Neufeld2019}, all of which rely upon the correspondence of each nonzero element to a single or double excitation. Here, multi-indices are used to denote these excitations. For example, $(K, 1, i, a)$ denotes a single excitation from $| K \rangle$ involving occupied orbital $i$ and virtual orbital $a$.

In this study, $\mathbf{P}_\text{off-diag}$ is factored according to the heat-bath Power-Pitzer (HB-PP) factorization~\cite{Holmes2016, Neufeld2019}. The following presentation of this scheme yields an algorithm equivalent to that used in our previous work, even though here $\mathbf{P}^{(\tau)}$ is partitioned into a sum of two matrices. Additional comments on these differences can be found in Appendix \ref{sec:HBPP}. The matrix $\mathbf{P}_\text{off-diag}$ is represented as the product
$\mathbf{B Q}$, where $\mathbf{Q}$ is the product of five matrices, $\mathbf{Q}^{(5)} \mathbf{Q}^{(4)} \mathbf{Q}^{(3)} \mathbf{Q}^{(2)} \mathbf{Q}^{(1)}$. Elements of $\mathbf{Q}$ correspond to the ``excitation generation probabilities'' used in FCIQMC, and its row space is indexed by single and double excitations. The row spaces of intermediate matrix factors of $\mathbf{Q}$ are smaller and indexed by only a subset of orbitals involved in each excitation, e.g. $(L, 2, i, j, a)$. 

Different excitations in the row space of $\mathbf{Q}$ can map to the same Slater determinant. For example, $(K, 1, i, a)$ and $(L, 2, i, j, a, b)$ both map to $| M \rangle$ if $| M \rangle = \left| \left( \hat{c}^\dagger_a \hat{c}_i | K \rangle \right) \right| = \left| \left( \hat{c}^\dagger_a \hat{c}^\dagger_b \hat{c}_i \hat{c}_j | L \rangle \right) \right|$. In the course of multiplication by $\mathbf{B}$, elements for excitations that map to the same determinant are summed. This step, together with the addition of the resulting vector to $\mathbf{P}^{(\tau)}_\text{diag} \Phi(\mathbf{v}^{(\tau)})$, is referred to as ``annihilation'' in the context of FCIQMC. More specifically, elements of $\mathbf{B}$ corresponding to single excitations are specified as
\begin{equation}
\label{eq:singB}
B_{M, (K, 1, i, a)} = \frac{P^{(\tau)}_{M, K}}{Q_{(K, 1, i, a), K}}
\end{equation}
and those for double excitations are specified as
\begin{multline}
\label{eq:doubB}
B_{M, (L, 2, i, j, a, b)} = P^{(\tau)}_{M, L} \left( Q_{(L, 2, i, j, a, b), L} + Q_{(L, 2, i, j, b, a), L} + \right. \\ \left. Q_{(L, 2, j, i, a, b), L} + Q_{(L, 2, j, i, b, a), L} \right)^{-1}
\end{multline}
where the determinant indices $K$, $L$, and $M$ are defined as in the example above. Four elements of $\mathbf{Q}$ are summed in eq \ref{eq:doubB} in order to account for the four different double excitations that map to each determinant.
All other elements of $\mathbf{B}$ are zero. 

Despite the reduced cost and memory requirements associated with this factorization scheme, performing these matrix-vector multiplications and compressing the resulting vectors constitute the cost and memory bottlenecks in our current implementation (although these steps can be parallelized). For many chemical systems, the steps involving vectors of length $\mathcal{O}(Vm)$ will limit the overall performance. This scaling could be improved by using more matrices in the factorization, each with fewer nonzero elements per column. Alternatively, segments of these vectors could be multiplied and compressed independently, thereby decreasing the degree of correlation enforced. FCIQMC methods invoke this strategy with a maximal degree of independence (minimal correlation), as segments corresponding to different walkers are treated entirely independently. The FRI framework admits compression schemes with an intermediate degree of independence, allowing for the possibility that the key correlations responsible for increased efficiency could be retained while reducing per-iteration costs. We leave the further optimization of FCI-FRI methods by way of this strategy to future studies.

\subsection{Estimating the Ground-State Eigenvalue and its Standard Error}
Having presented a method for generating stochastic iterates approximating the ground-state eigenvector, we next discuss how to use them to approximate the ground-state energy. In principle, one could average the iterates and calculate the energy of the resulting vector using the standard, variational Rayleigh quotient estimator. 
However, this would eliminate the memory advantages of the sparsity provided by the vector compression techniques described above, since it would require accumulating the average in a vector of the same dimension as $\mathbf{H}$.
The ground-state energy is therefore instead estimated as~\cite{Overy2014}
\begin{equation}
\label{eq:eMean}
\langle E \rangle = \frac{\sum_{\tau \geq \tau_c} n^{(\tau)}}{\sum_{\tau \geq \tau_c} d^{(\tau)}}
\end{equation}
where
\begin{equation}
\label{eq:estNumer}
n^{(\tau)} = \mathbf{v}^*_\text{ref} \mathbf{H v}^{(\tau)}
\end{equation}
and
\begin{equation}
\label{eq:estDenom}
d^{(\tau)} = \mathbf{v}^*_\text{ref} \mathbf{v}^{(\tau)}
\end{equation}
Here the sum excludes iterations with indices less than a burn-in time $\tau_c$, before which the values of $n^{(\tau)}$ and $d^{(\tau)}$ lie significantly outside the range of fluctuations observed later in the calculation. The vector $\mathbf{v}_\text{ref}$ is chosen as an approximation of the ground-state, usually calculated using an inexpensive electronic structure method. In this paper, for the sake of comparison, the Hartree-Fock unit vector is used as $\mathbf{v}_\text{ref}$, although less statistical error could be achieved by using a vector closer to the ground state.

This estimator does not formally converge after infinitely many iterations because the normalized ground-state eigenvector is determined only up to an arbitrary phase. For example, if the iterates are real and the sampling is ergodic, then an infinite-length calculation will have equal numbers of iterates with positive and negative signs (as determined by the sign of their inner product with an arbitrary vector). Averaging these iterates will yield the zero vector, so any quantity that depends linearly on the iterates, i.e. $n^{(\tau)}$ and $d^{(\tau)}$, will also average to zero~\cite{Vigor2016}. This can be rectified by fixing the signs of all iterates to be either positive or negative. During calculations of typical length ($\sim$1 million iterations), we found that iterates changed sign only when few nonzero elements were retained in compression operations, in which case the statistical error in the ground-state energy was on the order of 0.1 to 10 $E_h$. Nevertheless, we used this constraint in all calculations presented here because it can be applied as an inexpensive post-processing operation.
Even with this constraint, the average of the iterates does not converge to the exact ground-state eigenvector after infinitely many iterations. Although the compression operations are unbiased (eq \ref{eq:compDef}), elements in each matrix $\mathbf{P}^{(\tau)}$ 
depend on a quotient of correlated random numbers (eq \ref{eq:shiftUpdate}), which causes the iterates to be biased. This is often called the ``population control bias'' in QMC~\cite{Vigor2015}. In typical calculations, the magnitude of this bias is often less than the standard error.

The standard error $\sigma_e$ associated with the mean ground-state energy (eq \ref{eq:eMean}) is estimated as described in ref \citenum{Greene2019}
by applying standard Monte Carlo error estimation techniques to the sequence
\begin{equation}
\frac{n^{(\tau)}}{\langle d \rangle} - \frac{\langle n \rangle d^{(\tau)}}{\langle d \rangle^2}
\end{equation}
where $\langle n \rangle$ and $\langle d \rangle$ represent the trajectory means of $n^{(\tau)}$ and $d^{(\tau)}$, respectively. We use the emcee software package~\cite{Foreman2013} to estimate errors. The standard error decreases asymptotically as $N_i^{-1/2}$~\cite{Chung1960, Sokal1997}, where $N_i$ is the number of iterations included in the trajectory averages. The \textit{statistical efficiency} $E$ is therefore used as an error metric that is asymptotically independent of trajectory length:
\begin{equation}
\label{eq:eff}
E= \sigma_e^{-2} N_i^{-1}
\end{equation}
A greater statistical efficiency indicates a smaller standard error after a fixed number of iterations.

\subsection{Implementing FCI-FRI in Parallel}
\label{sec:code}
The computational efficiency of the systematic FCI-FRI method depends critically on the reduced CPU and memory costs associated with representing only the nonzero elements in the matrices and vectors in each iteration. There are many systems for which the ground-state FCI energy can be reliably estimated using significantly fewer nonzero elements in each iteration than the dimension of $\mathbf{H}$. However, as will be discussed below, there is often a lower limit to the number of nonzero elements needed to achieve a reliable estimate in a reasonable number of iterations. Because this number can become large as system size increases~\cite{Cleland2011}, it can be advantageous to distribute the elements among many parallel processes, e.g. by using the MPI framework. This section discusses some of the considerations involved in implementing the systematic FCI-FRI method in parallel. The source code for our implementation, written in C++, is freely available on GitHub~\cite{fries}.

A key requirement of any implementation that uses sparse vectors is the ability to efficiently query the value of an element at an arbitrary index. In our implementation, this is accomplished using the hashing techniques described in ref \citenum{Booth2014}. Briefly, a hash function is used to map each Slater determinant index to an MPI process, and a separate hash table within that process is used to locate the corresponding element. Thus, each element added to a vector in the course of matrix-vector multiplication can be added at $\mathcal{O}(1)$ cost.


Since vector elements are distributed among multiple processes, applying the compression scheme described in Section \ref{sec:vecComp} requires communication among processes. In order to ensure efficiency, the amount of information communicated should be minimized. The first step of this scheme involves locating the largest-magnitude vector elements in the subspace $\mathcal{D}$ according to the condition in eq \ref{eq:rhoCriterion}. One way to do this is by first calculating the sum of the magnitudes of all elements in parallel, then subtracting the magnitude of the largest element, then the second-largest, etc. until the condition is satisfied. These elements can be found efficiently by heaping the elements on each process independently, in parallel, communicating and comparing only the largest-magnitude element from each process, and updating the heaps as elements are removed from consideration. In practice, we have found that an iterative technique that leverages memory locality can be made more efficient when compressing the vectors resulting from multiplication by the matrix factors introduced in Section \ref{sec:Hfac}. This technique is based on the observation that elements in $\mathcal{D}$ can be selected in any order, as long as the criterion in eq \ref{eq:rhoCriterion} is checked for all remaining elements.  Further details can be found in our source code. Performing the second step of compression, in which the remaining nonzero elements are selected randomly, requires communicating only a single random number and the sums of magnitudes of the elements on each process.

\subsection{Chemical Systems for Numerical Tests}
\label{sec:systems}
The remainder of this paper describes comparisons among calculations of the ground-state energies for three systems: the Ne atom and the \ce{H2O} and \ce{N2} molecules.
The single-particle basis sets used for each system are reported in Table \ref{tab:molSystems}. The dimension of the corresponding FCI matrices, which depend combinatorially on the size of the single-particle basis ($M$) and the number of electrons ($N$) are also given in Table \ref{tab:molSystems}. Calculations were performed at the equilibrium geometries of the \ce{H2O} and \ce{N2} molecules, as reported in refs \citenum{Olsen1996a} and \citenum{Olsen1996}, respectively. In contrast to our previous work~\cite{Greene2019}, the single-particle basis sets used here were not truncated according to natural orbital occupancies.
These systems, which have been used previously to benchmark FCIQMC calculations~\cite{Booth2009}, are small enough that deterministic FCI results are available but large enough that poor convergence is observed when too few samples are used in compression operations. A detailed account of the input parameters for our code needed to reproduce all results presented in this manuscript is available in our GitHub repository~\cite{fries}.

\begin{table}
\caption{The parameters defining the FCI Hamiltonian matrix for each of the systems considered in this study. The parameter $N$ denotes the number of active electrons considered in each system (core electrons were frozen for Ne and \ce{N2}), $M$ is the number of active orbitals in the single-particle basis for each system, and $N_\text{FCI}$ is the total dimension of the ground-state symmetry block of $\mathbf{H}$. The ground-state energy, $E_\text{FCI}$, includes the nuclear repulsion energy.}
\begin{tabular}{c | c | c | c | c}
& Single-particle & & & \\
System & basis & $(N, M)$ & $N_\text{FCI} / 10^6$ &  $E_\text{FCI} / E_h$ \\ \hline
Ne & aug-cc-pVDZ & $(8, 22)$ & 6.69 &  $-128.709476^\text{a}$\\
\ce{H2O} & cc-pVDZ & $(10, 24)$ & 451 & $-76.241860^\text{b}$\\
\ce{N2} & cc-pVDZ & $(10, 26)$ & 541 &  $-109.276527^\text{a}$ 
\end{tabular}

\raggedright
$^\text{a}$From ref \citenum{Olsen1996}\\
$^\text{b}$From ref \citenum{Olsen1996a}
\label{tab:molSystems}
\end{table}

\section{The Initiator Approximation}
\label{sec:initiator}
The systematic FCI-FRI method can in principle approximate the ground-state eigenvalue of any FCI Hamiltonian matrix. Its performance for several small chemical systems was evaluated previously, in ref \citenum{Greene2019}. It is well-known that the convergence properties of FCIQMC can depend very strongly on the number of walkers when that number is small~\cite{Booth2011, Spencer2012, Kolodrubetz2013, Shepherd2014, Vigor2016}. Larger systems generally require more walkers, which can make it expensive to reliably estimate their energies. The initiator approximation~\cite{Cleland2010, Booth2011} was introduced in the FCIQMC context to address this issue of poor convergence. Our numerical tests indicate that the systematic FCI-FRI method exhibits similar behavior when few nonzero elements are used in compression. This section demonstrates that the initiator approximation can be applied straightforwardly to FCI-FRI methods in order to address the convergence issue.

Applying the initiator approximation involves replacing the matrix $\mathbf{B}$ in the factorization described in Section \ref{sec:Hfac} with a modified matrix $\mathbf{B}^{\prime (\tau)}$, in which some elements are zeroed in each iteration. Elements of $\mathbf{B}^{\prime (\tau)}$ are given as
\begin{equation}
{B}^{\prime (\tau)}_{L, (K, 1, i, a)} = 
\begin{cases}
0 & v^{(\tau)}_L = 0 \text{ and }\left| v^{(\tau)}_K \right| < n_a \\
B_{L, (K, 1, i, a)} & \text{ otherwise}
\end{cases}
\end{equation}
and
\begin{equation}
B^{\prime (\tau)}_{L, (K, 2, i, j, a, b)} = 
\begin{cases}
0 & v^{(\tau)}_L = 0 \text{ and }\left| v^{(\tau)}_K \right| < n_a \\
B_{L, (K, 2, i, j, a, b)} & \text{ otherwise}
\end{cases}
\end{equation}
where $n_a$ is the \textit{initiator threshold}~\cite{Cleland2011}. This approximation was designed to minimize contributions to the next iterate from elements in the current iterate with signs that are not well-established~\cite{Cleland2010}. Elements with magnitudes less than $n_a$, which are more likely to change sign in subsequent iterations, are prevented (via the zeroes introduced) from contributing weight to elements that are zero in the current iteration. Early implementations of initiator FCIQMC included an additional modification to $\mathbf{B}$ based on ``sign-coherent'' spawning events~\cite{Cleland2010}. It was later found that this additional rule makes little difference in practice~\cite{Blunt2015}, so we do not consider it here.

Although stochastic compression as used in FRI also involves zeroing elements, there is an important difference in how elements are zeroed in the initiator approximation. In stochastic compression, each nonzero element is zeroed with a probability less than 1, in order to ensure that its expected value equals its original value before compression (eq \ref{eq:compDef}). In the initiator approximation, some elements are zeroed with probability 1, meaning that their original value is not preserved in expectation. As in initiator FCIQMC, this introduces an additional source of bias in the eigenvector obtained using power iteration that is not present when the initiator approximation is not used. This bias is expected to decrease with increasing iterate norms, since fewer elements will have magnitudes less than $n_a$, and thus $\mathbf{B}^{\prime (\tau)}$ will converge to $\mathbf{B}$. Due to this implicit dependence of elements of $\mathbf{B}^{\prime (\tau)}$ on the norm of each iterate, all comparisons in this paper are performed using the same target one-norm (defined in Appendix \ref{sec:powMeth}), unless otherwise noted. When $n_a = 0$, this method yields the same results as when the initiator approximation is not applied.

\begin{figure}
\includegraphics[width=\linewidth]{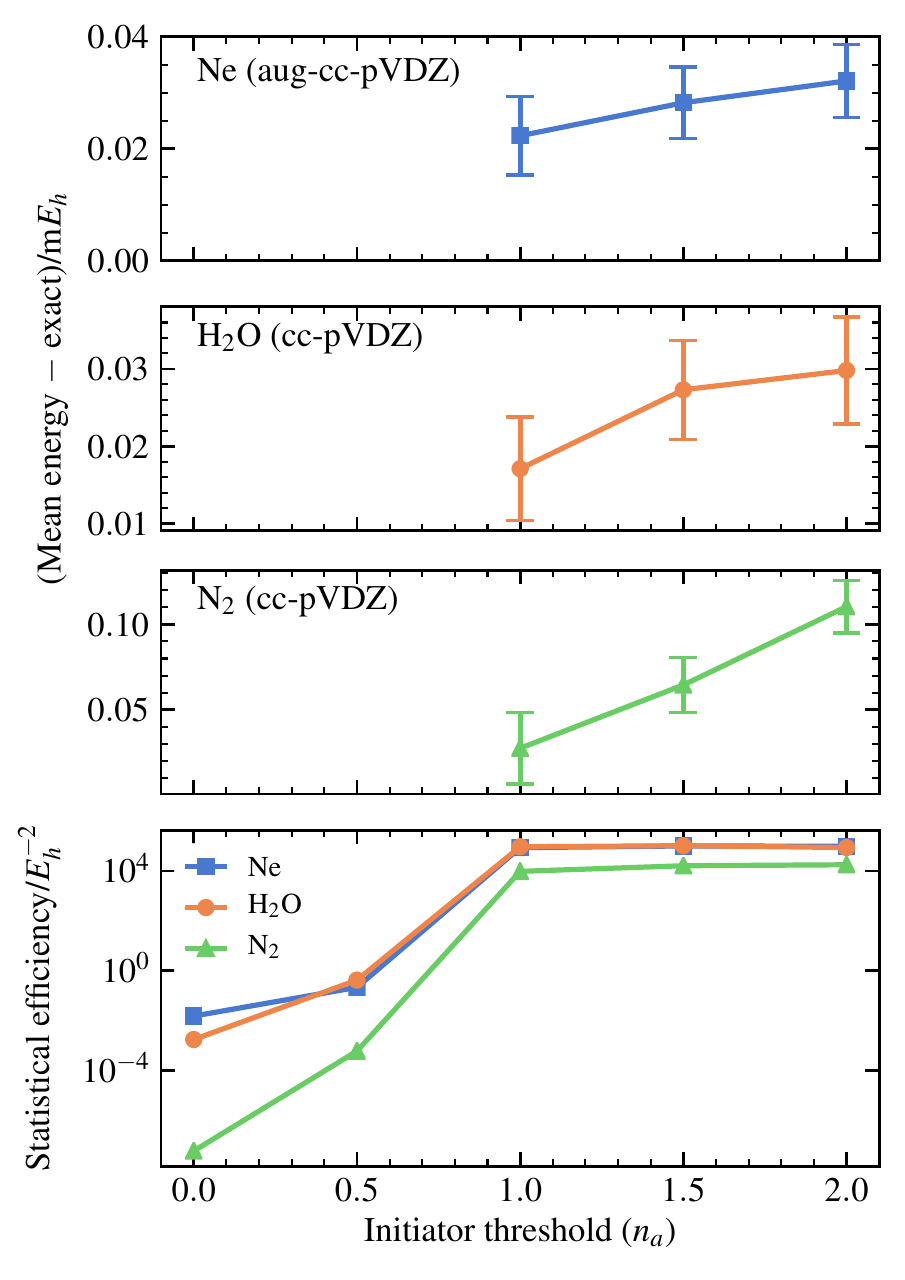}
\caption{(top) Ground-state energies for Ne, \ce{H2O}, and \ce{N2}, estimated using the FCI-FRI method with different values of the initiator threshold, $n_a$. For Ne, vectors were compressed to 100,000 nonzero elements, and those for \ce{H2O} and \ce{N2} were compressed to 1 million. The exact ground-state energy for each system is subtracted from the estimate. Error bars indicate 95\% confidence intervals ($\pm 2\sigma_e$) after 1 million iterations. Estimates for the first two values of $n_a$ (0 and 0.5) are not shown, since their standard errors greatly exceed the range of the vertical axis. (bottom) The statistical efficiency associated with each estimate, calculated according to eq \ref{eq:eff}. Note the dramatic increase in statistical efficiency for all systems as $n_a$ is increased from 0 to 1.}
\label{fig:iniThresh}
\end{figure}

We applied the FCI-FRI method with the initiator approximation to the three systems described in Section \ref{sec:systems}; results are presented in Figure \ref{fig:iniThresh}. For Ne, the vector obtained after each matrix multiplication in the Hamiltonian matrix factorization was compressed to $m=100,000$ nonzero elements. For \ce{H2O} and \ce{N2}, vectors were compressed to $m=1$ million elements. For all three systems, the target one-norm was specified as $m$. The statistical efficiencies of these calculations, and correspondingly their standard errors after 1 million iterations, depend strongly on the initiator threshold, $n_a$, for $n_a \leq 1$. Standard errors at $n_a = 0$ are on the order of $0.01 - 10~E_h$, i.e. significantly outside the range acceptable for chemical accuracy. This demonstrates that it is difficult to obtain a good estimate of the ground-state energy after a reasonable number of iterations when too few elements are used in vector compression. For calculations with $n_a \geq 1$, the standard error is sufficiently small to ascertain that the mean energy differs significantly from the exact energy. This statistical bias originates from two sources: (1) the bias inherent in all stochastic implementations of the power method, and (2) the additional bias introduced by the initiator approximation. As $n_a$ is increased beyond 1, the bias increases for Ne and \ce{N2}, while the statistical efficiency for all systems does not change appreciably. Similar behavior has been observed previously in the context of FCIQMC~\cite{Shepherd2012a}, where it was suggested that increasing $n_a$ limits increases in statistical efficiency that would otherwise be expected by slowing the transfer of weight among elements in the solution vector. Our results suggest that $n_a = 1$ is a suitable choice of the initiator threshold for the FCI-FRI method, at least for these systems. A statistically significant bias of $0.17 \pm 0.04~E_h$. was also observed in the \ce{H2O} calculation with $n_a = 0$, and a bias of $0.28 \pm 0.08~E_h$ was observed for \ce{N2} at $n_a = 0.5$.

\begin{figure}
\includegraphics[width=\linewidth]{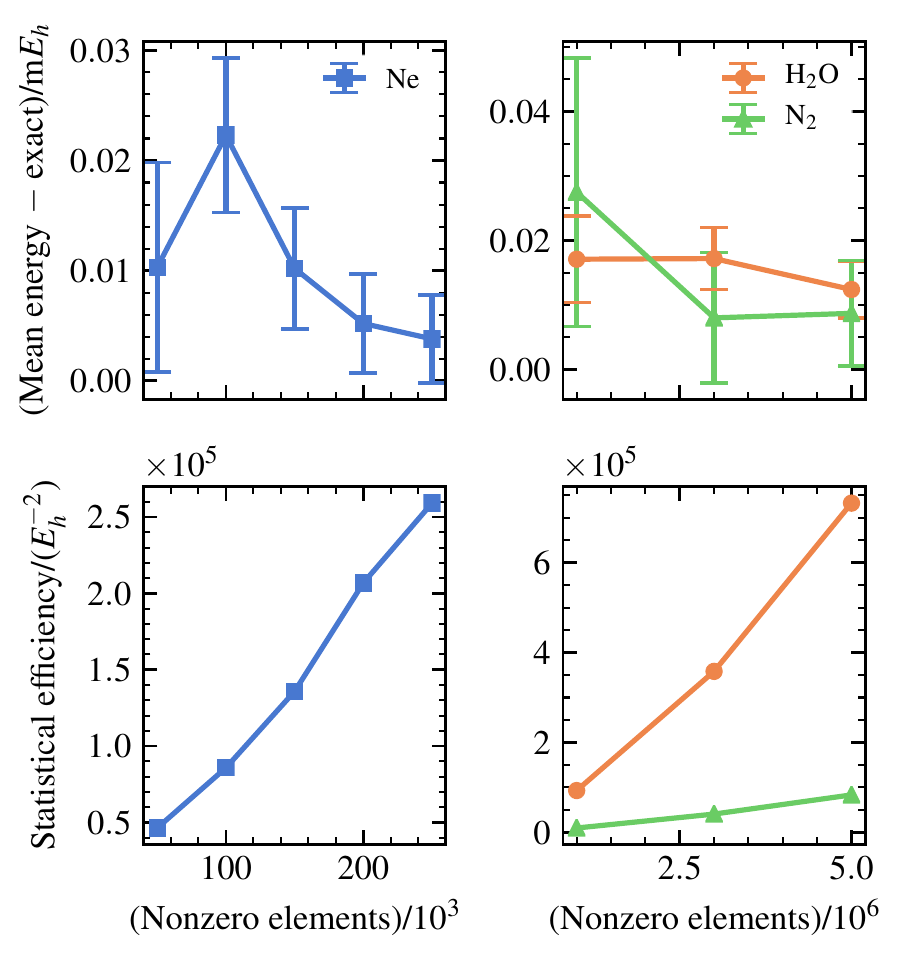}
\caption{(top) Ground-state energies as calculated using the FCI-FRI method with an initiator threshold of $n_a = 1$. The horizontal axis indicates the target number of nonzero elements $(m)$ used for all compression operations in each iteration. Error bars indicate 95\% confidence intervals ($\pm 2 \sigma_e$). (bottom) The statistical efficiency for each calculation.}
\label{fig:iniNonzTrend}
\end{figure}

Figure \ref{fig:iniNonzTrend} shows the behavior of the initiator FCI-FRI method as the number of nonzero elements used in each compression $(m)$ is increased. An initiator threshold of $n_a = 1$ was used in all calculations, and the target one-norm was specified as $m$. For all systems, the statistical efficiency increases approximately linearly with $m$. This indicates a $m^{-1/2}$ dependence of the standard error, as is expected for $m$ sufficiently large. The results suggest that the bias depends weakly on $m$, although it is difficult to draw conclusions with confidence due to the magnitude of the errors. It may be possible to reduce the bias as $m$ is increased by decreasing the initiator threshold $n_a$ or, equivalently, increasing the target one-norm, since $n_a$ can be set to 0 for sufficiently large $m$. We leave the investigation of this possibility for future studies.

\section{Semi-stochastic FCI-FRI}
\label{sec:semiStoch}
The FCI-FRI methods described above use the criterion specified in eq \ref{eq:rhoCriterion} to dynamically select a subspace $\mathcal{D}$ that contains elements to be preserved exactly in each compression operation. This section discusses the potential benefits of constraining $\mathcal{D}$ to contain both a fixed set $\mathcal{D}_\text{fixed}$ of elements that are preserved exactly, regardless of their magnitudes, and a dynamic set $\mathcal{D}_\text{dynam}$ chosen as described above. For clarity in the following presentation, we specify that $\mathcal{D}_\text{fixed} \cup \mathcal{D}_\text{dynam} = \mathcal{D}$ and $\mathcal{D}_\text{fixed} \cap \mathcal{D}_\text{dynam} = \emptyset$. This modification to the FCI-FRI method was motivated by the semi-stochastic extension to FCIQMC (s-FCIQMC)~\cite{Petruzielo2012, Blunt2015b}. 

Before describing possible choices of $\mathcal{D}_\text{fixed}$, we will first describe how to perform compression given a particular choice of $\mathcal{D}_\text{fixed}$. In our implementation, the two-step compression algorithm described in Section \ref{sec:vecComp} is applied only to the elements \textbf{not} in $\mathcal{D}_\text{fixed}$, and fewer than $m$ nonzero elements are selected from among these elements. In the first step, the number of elements in $\mathcal{D}_\text{dynam}$ is determined according to a modified version of the criterion in eq \ref{eq:rhoCriterion}. If $\mathbf{x}$ is the vector being compressed, $\mathcal{D}_\text{dynam}$ contains the $\rho$ largest magnitude elements not in $\mathcal{D}_\text{fixed}$, where $\rho$ is the minimum value of $h$ for which
\begin{equation}
\label{eq:semistochComp}
(m - d - h) |x_{s_{h+1}}| \leq \sum_{j=h+1}^{||\mathbf{x}||_0 - N_\text{determ}} |x_{s_j}|
\end{equation}
where $d$ is defined in the following paragraph, and the vector $\mathbf{s}$ sorts only the elements of the input vector $\mathbf{x}$ not in $\mathcal{D}_\text{fixed}$. The number of elements in $\mathbf{x}$ in $\mathcal{D}_\text{fixed}$ is denoted as $N_\text{determ}$. After determining $\mathcal{D}_\text{dynam}$, $(m - d - \rho)$ nonzero elements are sampled from the set of elements not in $\mathcal{D}$. As mentioned in Section \ref{sec:vecComp}, choosing $\mathcal{D}$ to include a different set of elements than indicated by the original criterion in eq \ref{eq:rhoCriterion} is sub-optimal in terms of the statistical error incurred in a single compression operation. However, other choices of $\mathcal{D}$, such as the one described in this section, can potentially yield less statistical error in the broader context of the stochastic power method. For example, it can be advantageous to exactly preserve elements that have large magnitudes in the exact eigenvector, even if their magnitudes in the current iterate are small.

In semi-stochastic FCI-FRI, this compression scheme is applied to each of the vectors obtained after multiplication by the Hamiltonian matrix factors discussed in Section \ref{sec:Hfac}. Since these vectors have different dimensions, a brief discussion of how we specify $\mathcal{D}_\text{fixed}$ for each vector is warranted. Iterates exist in a space of Slater determinants, with dimension $N_\text{FCI}$, whereas vectors obtained after multiplication by the matrices comprising $\mathbf{Q}$ exist in spaces of excitations from determinants, with dimensions greater than $N_\text{FCI}$. Bases for excitations are indexed by multi-indices, which in all cases contain a Slater determinant as the first component. Here, the sets $\mathcal{D}_\text{fixed}$ for all vectors within a single iteration are specified by Slater determinants. In the case of elements indexed by excitations, these correspond to the first component in their respective multi-indices. In other words, if the Slater determinant index $K$ is in this fixed set, then the multi-indices $(K, 1, i, a)$ and $(K, 2, i, j, a, b)$ for all $i,j,a,$ and $b$ are in the sets $\mathcal{D}_\text{fixed}$ for their respective vectors. The value of $d$ in eq \ref{eq:semistochComp} used in each compression is chosen to facilitate comparison to calculations in which the semi-stochastic extension is not used. The same value of $d$ is used for compressing the vectors resulting from multiplication by each of the matrix factors of $\mathbf{Q}$: it is the number of nonzero Hamiltonian matrix elements corresponding to excitations from determinants in $\mathcal{D}_\text{fixed}$. This choice of $d$ ensures that the number of Hamiltonian elements evaluated is the same as in calculations without the semi-stochastic extension, and it obviates the need to explicitly enumerate or count the elements in $\mathcal{D}_\text{fixed}$ for these vectors. For the compression operation preceding multiplication by the first matrix factor of $\mathbf{Q}$, $d$ is simply the number of determinants in $\mathcal{D}_\text{fixed}$.

\begin{figure*}
\includegraphics[width=\linewidth]{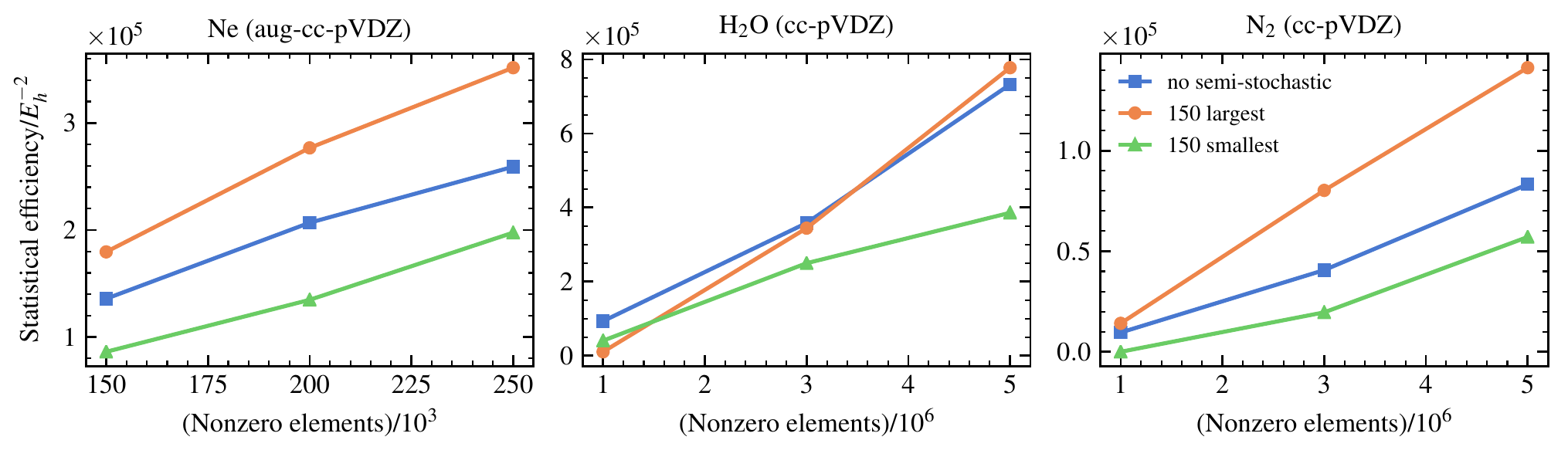}
\caption{A comparison of the statistical efficiencies of FCI-FRI calculations performed with and without the semi-stochastic extension. Two sets of semi-stochastic calculations were performed for each system: one using a subspace consisting of the largest-magnitude elements from the CISD ground-state eigenvector, and the other using the smallest elements. The calculations performed without the semi-stochastic extension correspond to those shown in Figure \ref{fig:iniNonzTrend}. All calculations at common values of the number of nonzero elements ($m$) yielded mean energies that agreed to within statistical uncertainty ($2 \sigma_e$).}
\label{fig:semistoch}
\end{figure*}

This specification of the deterministic subspace differs from the one typically used in s-FCIQMC. In effect, entire columns of the Hamiltonian matrix corresponding to Slater determinants in the fixed deterministic subspace are left unchanged after these compression operations, whereas in s-FCIQMC Hamiltonian elements are only preserved exactly if they connect two determinants in the deterministic subspace. The approach described here offers several advantages in the context of FCI-FRI. To our knowledge, a compression scheme for FCIQMC that excludes elements in the deterministic part of the Hamiltonian does not exist. Thus, in implementations of s-FCIQMC, excitations from the deterministic subspace are included in compression operations~\cite{Blunt2015b}. The matrix $\mathbf{B}$ is modified such that elements within the deterministic block of the Hamiltonian are zero, and this block is multiplied separately to compensate. In our implementation, excitations from the deterministic subspace are not included in compression operations, which reduces their cost. Additionally, this approach avoids the statistical error incurred in compression operations in s-FCIQMC by including Hamiltonian elements that couple determinants within the deterministic subspace to those outside.

We will next discuss the considerations involved in choosing the determinants to include in the fixed subspace. In early implementations of s-FCIQMC~\cite{Petruzielo2012}, they were chosen as the greatest-magnitude elements in the ground-state eigenvector obtained by diagonalizing $\mathbf{H}$ in a larger subspace. This larger subspace was constructed by repeatedly applying $\mathbf{H}$ to a trial vector and truncating deterministically, as is done similarly in selected configuration interaction methods~\cite{Holmes2016a, Sharma2017}. In current implementations of s-FCIQMC, the fixed subspace is specified as that containing the largest-magnitude elements from the last iterate of a preliminary s-FCIQMC calculation executed with a simple fixed subspace, e.g. the space of single and double excitations from Hartree-Fock (CISD)~\cite{Blunt2015b}. Since this section is intended to serve only as a preliminary exploration of the potential benefits (or downsides) of applying the semi-stochastic extension to FCI-FRI, we compare two simple choices of the fixed subspace, which are designed to represent a ``good'' choice and a ``bad'' choice. The ``good'' choice contains the largest-magnitude elements from the CISD ground state and the ``bad'' choice contains the smallest-magnitude elements. We leave the development and implementation of subspace selection methods for FCI-FRI to future studies.

The results of this comparison are presented in Figure \ref{fig:semistoch}. All calculations were performed with the target one-norm fixed at the number of elements used in each compression operation ($m$) and an initiator threshold of $n_a = 1$. Calculations with three different values of $m$ were performed for each system. At each value of $m$, we compare the statistical efficiency of three calculations: two that use each of the choices of the fixed deterministic subspace described above, and one without the semi-stochastic extension. The fixed deterministic subspaces contained 50 Slater determinants for Ne, and 150 for \ce{H2O} and \ce{N2}. The number of nonzero Hamiltonian matrix elements corresponding to excitations from these determinants is approximately 49,000 for Ne, 509,000 for \ce{H2O}, and 311,000 for \ce{N2}, which determines the values of $d$ used in compression operations after multiplication by the Hamiltonian matrix factors. Because the total number of matrix elements evaluated in each iteration is fixed, the cost of calculations with and without the semi-stochastic extension were approximately the same. 

For all three systems, using a deterministic subspace defined by the smallest-magnitude determinants from the CISD eigenvector reduced the statistical efficiency relative to the calculation without the semi-stochastic extension, by as much as six orders of magnitude for the \ce{N2} calculation with $m = 1$ million. Instead using the largest-magnitude CISD determinants increased the statistical efficiencies for Ne and \ce{N2} calculations, by at most a factor of 2. Trends in statistical efficiency for \ce{H2O} calculations are less clear. Using the ``good'' deterministic subspace \textit{reduced} the statistical efficiency for calculations with $m=1$ million and only marginally increased it for $m = 5$ million. Together, these results suggest that the criteria for choosing a good deterministic subspace may depend on the system under consideration, and that the semi-stochastic extension does not always improve the performance of the systematic FCI-FRI method when the number of matrix element evaluations is fixed.

\section{An Alternative Hamiltonian Matrix Factorization}
\label{sec:newHB}
The calculations described thus far use the heat-bath Power-Pitzer (HB-PP) factorization to perform the matrix-vector multiplication in each iteration at an affordable computational cost. In this section, we show that FCI-FRI methods can achieve better performance by using a modified form of this factorization. We begin by explaining the motivation behind these modifications.

The matrix factors comprising the matrix $\mathbf{Q}$, defined in Section \ref{sec:Hfac}, share two features that follow from the development of the HB-PP factorization for FCIQMC: (1) there is only one nonzero element in each row, and (2) the elements in each column are positive and sum to 1. Due to the first feature, the row spaces of these matrices can be divided into disjoint subspaces, each corresponding to a single element in the vector being multiplied. This facilitates the straightforward parallelization of \textit{stratified} compression techniques used in FCIQMC, in which these subspaces are treated independently and in parallel. This sequence of stratified compression operations can be formulated as the random selection of each of the orbitals comprising excitations from Slater determinants. The elements in each column correspond to probabilities for each orbital, which dictates the constraint specified in the second feature. Due to the prohibitive memory requirements of storing probabilities for each Slater determinant in the basis, probabilities are calculated on-the-fly from the one- and two-electron integrals defining the Hamiltonian. In our implementation, the calculation of these normalizing factors for each column constitutes a significant fraction of the overall computational effort for FCI-FRI simulations.

A key feature of the FCI-FRI compression scheme described in Section \ref{sec:vecComp} that enables reductions in statistical error is that elements are not treated independently. Parallelization is achieved by other means, as discussed in Section \ref{sec:code}. Stratification is therefore not used in systematic FCI-FRI, so the constraint of normalized columns provides no advantages. Removing this constraint in FCI-FRI methods affords a potential advantage beyond the reduced computational cost associated with not calculating normalizing factors. Less statistical error is achieved when elements of the matrix $\mathbf{B}$, which depend on elements in $\mathbf{P}$ and $\mathbf{Q}$ (eqs \ref{eq:singB} and \ref{eq:doubB}), are more uniform in magnitude~\cite{Neufeld2019}. In an effort to reduce the variability in these magnitudes, we developed an alternative HB-PP factorization, in which the column norms of $\mathbf{Q}$ vary approximately in proportion to those of $\mathbf{P}$. An additional feature of this new factorization is that an ordering is enforced among the orbitals specifying double excitations. This reduces the row dimension of $\mathbf{Q}$, enabling vector compression to be performed with less statistical error. Thus, double excitation elements of the matrix $\mathbf{B}$ in this new factorization are given as (cf. eq \ref{eq:doubB})
\begin{equation}
B_{M, (L, 2, i, j, a, b)} = \frac{P^{(\tau)}_{M, L}}{Q_{(L, 2, i, j, a, b), L}}
\end{equation}
for $| M \rangle = \left| \left( \hat{c}^\dagger_a \hat{c}^\dagger_b \hat{c}_i \hat{c}_j | L \rangle \right) \right|$. The definitions of double excitation elements in this modified $\mathbf{Q}$ matrix are provided in Appendix \ref{sec:HBPP}. A variety of other techniques for reducing the variability in elements of $\mathbf{B}$ have been investigated previously~\cite{Neufeld2019}. Although we do not consider them here, many can be straightforwardly adapted for use in FCI-FRI and incorporated into this alternative factorization scheme.

\begin{figure*}
\includegraphics[width=\linewidth]{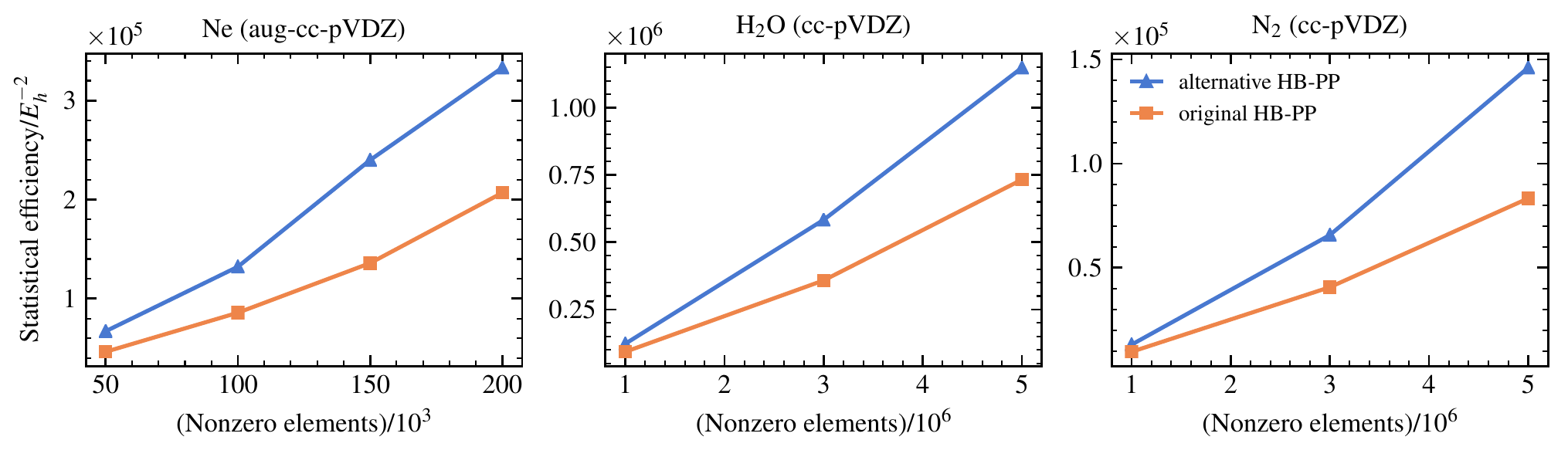}
\caption{The statistical efficiencies for calculations that use the original HB-PP Hamiltonian matrix factorization, in comparison to those that use the alternative factorization proposed here. The initiator approximation with threshold $n_a=1$ was used for all calculations, and the semi-stochastic extension was not used.}
\label{fig:newHB}
\end{figure*}

A direct comparison of the statistical efficiencies of calculations performed using the original factorization versus the alternative version is presented in Figure \ref{fig:newHB}. At each value of the number of nonzero elements used in compression operations ($m$), the calculation that uses the alternative factorization has a greater statistical efficiency. The relative advantage of using the alternative factorization, as measured by the ratio of statistical efficiencies, increases with $m$ for all three systems considered here. The greatest advantage was observed for Ne at $m = 150,000$, with a ratio of 1.8. The overall computational cost of calculations that use the alternative factorization was also 17\% \textit{less} than those that used the original factorization, on average. The biases for the \ce{N2} calculations with $m = 1$ million and $m = 3$ million ($0.070 \pm 0.018$ m$E_\text{h}$ and $0.026 \pm 0.008$ m$E_\text{h}$, respectively) are slightly greater than in calculations that used the original factorization. These observations exemplify the utility of the generic FRI framework for developing improvements whose benefits might be less apparent in a walker-based framework like FCIQMC. 
We elaborate further on comparisons between FCI-FRI and FCIQMC in the next section.

\section{Comparison to Initiator FCIQMC}
\label{sec:frivsFCIQMC}
Although the similarities between FCI-FRI and FCIQMC facilitate the application of the modifications discussed in the previous sections, the differences between the methods have implications for their relative performance. The primary difference is the degree of independence enforced in compression operations. In the first step of the vector compression scheme used in systematic FCI-FRI, a subset of elements are preserved exactly based on their relative magnitudes. Some FCIQMC implementations also allow for exact preservation of elements with magnitudes greater than a specified threshold~\cite{Blunt2015b}, but the key difference is that whether any one element is preserved exactly in FCIQMC is independent of whether any other element is preserved exactly. Additionally, the random selection of nonzero elements during the second step of the systematic FCI-FRI compression scheme is correlated, i.e. whether any particular element is selected determines which other elements are selected. In contrast, the random selection of excitations from any one Slater determinant in FCIQMC is independent of the excitations sampled for other determinants. Another difference is that compression operations in the original implementation of FCIQMC~\cite{Booth2009} include an additional constraint requiring vector elements to be integers. This constraint was relaxed in later FCIQMC implementations~\cite{Blunt2015b}, which allow for non-integer (floating-point) walker weights.



\begin{figure*}
\includegraphics[width=\linewidth]{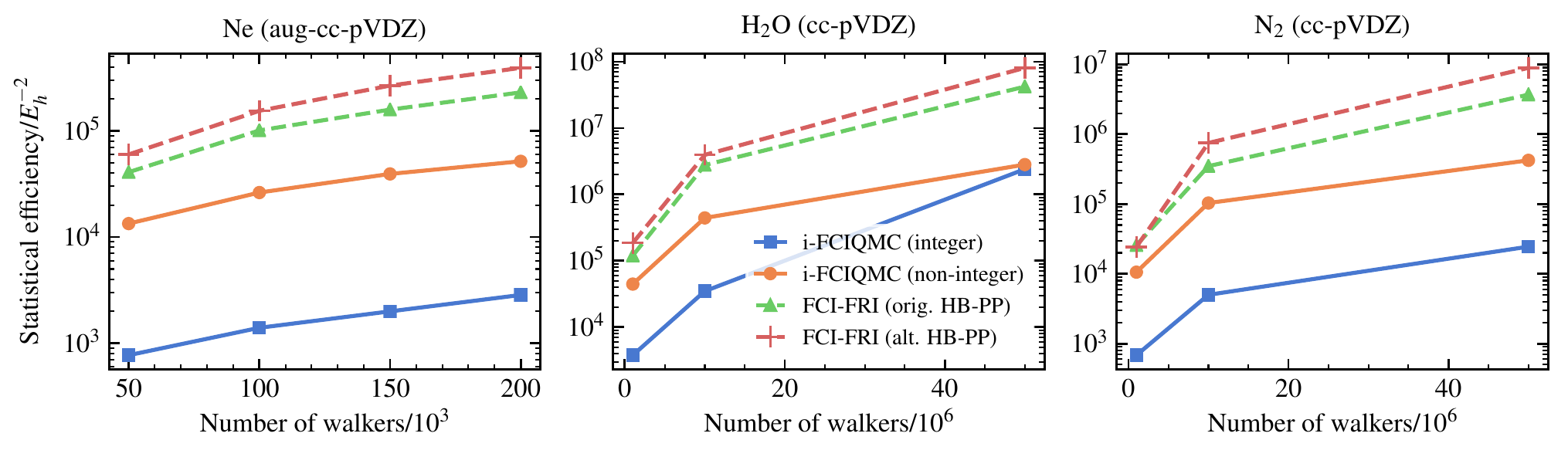}
\caption{Statistical efficiencies for systematic FCI-FRI and i-FCIQMC calculations executed with an initiator threshold of $n_a= 3$ for the Ne, \ce{H2O}, and \ce{N2} systems. In ``i-FCIQMC (integer)'' calculations, all Hamiltonian matrix elements are stochastically integerized {before} the ``annihilation'' step. In ``i-FCIQMC (non-integer)'' calculations, only a small subset of elements are integerized according to the method described in ref \citenum{Blunt2015b}. The horizontal axis denotes the number of walkers in i-FCIQMC calculations. The average distribution of walkers among Slater determinants determined the number of nonzero elements used to perform compressions in systematic FCI-FRI calculations, as described in the text. Systematic FCI-FRI results are presented for both versions of the HB-PP Hamiltonian matrix factorization discussed in this paper.}
\label{fig:fcimcVsFRI}
\end{figure*}

In Figure \ref{fig:fcimcVsFRI}, we compare results from FCI-FRI and two flavors of FCIQMC in order to quantify the effects of these differences on statistical efficiency. All calculations were performed with the initiator approximation, using a threshold of $n_a = 3$, but without any semi-stochastic extensions. Calculations were performed using our own implementations of these two FCIQMC algorithms~\cite{fries}. In the ``i-FCIQMC (integer)'' method, elements are constrained to be integers according to the procedure described in ref \citenum{Cleland2011}. The ``i-FCIQMC (non-integer)'' method corresponds to the method described in ref \citenum{Blunt2015b}, although it does not include the semi-stochastic extensions discussed therein. Only the elements with the smallest magnitudes are integerized in compression operations in order to reduce computational cost. The FCI-FRI (orig. HB-PP) method corresponds to the one presented in Section \ref{sec:initiator}, while the FCI-FRI (alt. HB-PP) calculations used the alternative Hamiltonian matrix factorization described in Section \ref{sec:newHB}.

Comparisons in Figure \ref{fig:fcimcVsFRI} are performed at a fixed walker number, corresponding to the target one-norm specified in FCIQMC calculations (Appendix \ref{sec:powMeth}). Since the FCI-FRI method is not formulated in terms of walkers, the target number of nonzero elements to use in each stochastic compression operation must be determined empirically from each FCIQMC calculation in order to enable these comparisons. In FCIQMC methods, the number of elements stochastically sampled from the Hamiltonian matrix in each iteration is the number of walkers. In the FCI-FRI context, this corresponds to the number of nonzero elements used in compression operations following multiplication by each of the matrix factors of $\mathbf{Q}$. Thus, the number of nonzero elements used in these compression operations in FCI-FRI was fixed at the average number of walkers in FCIQMC. 
The number of nonzero elements in FCIQMC iterates is determined by the distribution of walkers among Slater determinants. Therefore, in FCI-FRI, the target number of nonzero elements used in the compression operation preceding multiplication by the first matrix factor of $\mathbf{Q}$ was fixed at the average number of nonzero elements in FCIQMC iterates.
We emphasize that these constraints, which are enforced in order to enable a comparison between FCIQMC and FCI-FRI, represent suboptimal choices in FCI-FRI and that improved performance can be achieved by adjusting these parameters. For example, specifying the number of elements in the compression preceding multiplication by $\mathbf{Q}$ as the number of walkers instead of the number of nonzero iterate elements was found to increase the statistical efficiency more than the computational cost.

For all three chemical systems tested, mean energies from these three methods agree to within twice the standard error at each walker number. Statistical efficiencies from the systematic FCI-FRI method are 2.4 to 15 times greater than those from the ``i-FCIQMC (non-integer)'' method, which are in turn 1.2 to 21 times greater than those from the ``i-FCIQMC (integer)'' method. These results suggest that, although the use of non-integer elements in systematic FCI-FRI accounts for some of the gain in statistical efficiency relative to the ``i-FCIQMC (integer)'' method, the use of correlation also provides a consistent and significant improvement. We expect the parallel communication costs to be greater in the ``i-FCIQMC (non-integer)'' and FCI-FRI methods than in ``i-FCIQMC (integer).'' Using the non-integer implementation increases overall execution time by at most 50\% relative to the integer implementation, and FCI-FRI calculations were at most 41\% slower than ``i-FCIQMC (non-integer)'' calculations. Using the alternative HB-PP factorization in FCI-FRI calculations resulted in up to a 29-fold increase in statistical efficiency relative to the ``i-FCIQMC (non-integer)'' method.

All FCIQMC and FCI-FRI methods can be understood as a series of sequential matrix-vector multiplications and compression operations. The methods discussed here differ mainly in their approaches to compression. Our results indicate that enforcing correlations among elements, as is done in systematic FCI-FRI, improves statistical efficiency, albeit with somewhat increased cost and storage requirements relative to FCIQMC. In practice, one need not be confined to a choice solely between more expensive methods with correlations versus less expensive methods that treat elements independently. Within the basic FCI-FRI approach there are many possible methods with varying degrees of correlation and cost. Future research could involve investigating the trade-offs between cost and statistical error for different methods, and whether some methods are better suited to particular problems than others.

\section{Conclusions}
\label{sec:concl}
We demonstrated the applicability of three independent modifications to the FCI-FRI methods introduced in ref \citenum{Greene2019}. The initiator approximation was found to significantly improve performance when few nonzero elements are used in compression operations. Increasing the initiator threshold yielded consistent improvements in statistical efficiency up to a value of $n_a = 1$, but further increases yielded greater biases without significant improvements to the statistical efficiency. At a fixed initiator threshold, the statistical efficiency increases approximately in proportion to the number of nonzero elements used in calculations, while the bias remains constant (to within statistical uncertainty). The semi-stochastic extension with a good choice of deterministic subspace was found to improve the statistical efficiency for the Ne and \ce{N2} systems, but trends for \ce{H2O} were less clear. Our alternative heat-bath Power-Pitzer (HB-PP) matrix factorization was found to yield consistent improvements in statistical efficiency and reductions in computational cost for all systems tested.

These findings provide some insight into how the parameters in the systematic FCI-FRI method, in its present form, should be chosen to minimize cost and error. For example, it is advantageous to use the same number of nonzero elements in all compression operations in each iteration rather than the varying number used in our comparisons to FCIQMC. Additionally, our results above indicate clear benefits to using the initiator approximation and the modified HB-PP factorization. The inconsistent performance of the semi-stochastic extension observed in our tests suggests that further investigation is needed before we can recommend using it in FCI-FRI calculations.

The primary factor that determines the computational cost of FCI-FRI calculations is the number of nonzero elements used in compression operations. Previous studies suggest that the number of walkers required in FCIQMC scales weakly exponentially with system size~\cite{Cleland2011}. Due to the similarities between the two methods, we suspect that the required number of nonzero elements scales similarly in FCI-FRI. More research is needed to definitively determine whether this is the case.

In the results presented so far we have focused on benchmark applications for which full FCI results are available for comparison.  It is natural to ask whether and to what degree the improvements and prescriptions we describe here apply to more challenging systems, particularly those involving larger basis sets and stronger correlation. To provide some preliminary indication we performed several additional simulations which we describe now. We applied systematic FCI-FRI to the Ne atom in a cc-pVQZ basis. Using $m = 500,000$ nonzero elements in compression operations yielded a correlation energy estimate of $-333.41 \pm 0.017$ m$E_h$. This estimate was obtained using the alternative HB-PP factorization, which yielded a 2.5-fold increase in statistical efficiency relative to the original HB-PP factorization. It differs by approximately 1.3 m$E_h$ from the (non-initiator) FCIQMC estimate reported in ref \citenum{Booth2009}, obtained using 681 million walkers. We also applied systematic FCI-FRI to a more strongly correlated system, namely the \ce{N2} molecule at a stretched geometry in the cc-pVDZ basis. The resulting energy estimate obtained with $m = 5$ nonzero elements was within $0.10 \pm 0.10$ m$E_h$ of the exact FCI energy. Further computational details for these calculations are included in Appendix \ref{sec:hardParams}. As we pursue further improvements to FCI-FRI methods, we will continue to evaluate their applicability to these and other more challenging problems.

The scope of possible design features of randomized methods is broader than what has been tested previously in the context of either FCIQMC or FCI-FRI alone. In particular, many additional recent advances in FCIQMC methodology could be applied to FCI-FRI methods. We are currently exploring the possibility of including the unbiasing procedure from FCIQMC~\cite{Ghanem2019} and extending our methods for the calculation of properties other than the ground-state energy\cite{Booth2012, Blunt2014, Overy2014, Humeniuk2014, Blunt2015a, Blunt2015}. More generally, one could combine a variety of approaches to introducing independence and correlations in compression schemes, drawing upon ideas from FCIQMC and from the FRI framework, as a means of optimizing both computational cost and statistical error. A rigorous understanding of the advantages and disadvantages of each of these features can facilitate the development of generic FCI-FRI methods for treating strongly correlated systems beyond the capabilities of conventional quantum chemistry methods.

\appendix
\section{The Deterministic Power Method}
\label{sec:powMeth}
The sequence of power method iterates is defined by the relation
\begin{equation}
\mathbf{v}^{(\tau + 1)} = \mathbf{P}^{(\tau)} \mathbf{v}^{(\tau)}
\end{equation}
where 
\begin{equation}
\label{eq:Pdef}
\mathbf{P}^{(\tau)} = \mathbf{1} - \varepsilon \left(\mathbf{H} - S^{(\tau)} \mathbf{1} \right)
\end{equation}
This sequence converges to the ground-state eigenvector of $\mathbf{H}$ as $\tau \to \infty$, provided that $\varepsilon$ is sufficiently small and $\mathbf{v}^{(0)}$ is not orthogonal to the ground state. In this work, $\mathbf{v}^{(0)}$ is chosen as the ground-state eigenvector of the Hamiltonian projected into the space of all single and double excitations from the Hartree-Fock determinant (i.e. CISD).
The scalar-valued energy shift $S^{(\tau)}$ is included to stabilize the norms of the iterates. 
It is updated dynamically, at intervals of $A$ iterations, as follows:
\begin{equation}
\label{eq:shiftUpdate}
S^{(\tau)} = S^{(\tau - A)} - \frac{\xi}{A \varepsilon} \ln \frac{|| \mathbf{v}^{(\tau)}||_1}{|| \mathbf{v}^{(\tau - A)}||_1}
\end{equation}
where $\xi$ is a damping parameter used to reduce fluctuations in the shift, and $|| \cdot ||_1$ denotes the vector one-norm (i.e., the sum of the magnitudes of all elements). In this study, we use $A=10$ and $\xi = 0.05$, following previous studies~\cite{Booth2009}. In order to facilitate comparisons with FCIQMC calculations, the shift is fixed at 0 until the one-norm becomes greater than a target value. In the numerical tests presented here, the one-norm stabilizes at a value that is at most 10\% greater than the target.


\section{Modified Heat-Bath Power-Pitzer Factorization}
\label{sec:HBPP}
This section describes in more detail the alternative Hamiltonian matrix factorization scheme used to perform calculations in Section \ref{sec:newHB}. We provide formulas for elements of the five matrices whose product is the matrix $\mathbf{Q}$, defined in Section \ref{sec:Hfac}.

As in the original HB-PP factorization, a matrix $\mathbf{D}$ and vector $\mathbf{S}$ are calculated and stored at the beginning of each calculation. Each element of $\mathbf{D}$ and $\mathbf{S}$ approximates the sum of all Hamiltonian matrix elements corresponding to double excitations from a pair of occupied orbitals or a single orbital, respectively. Elements of $\mathbf{D}$ are calculated from the two-electron integrals from Hartree-Fock:
\begin{equation}
D_{pq} = \begin{cases}
\sum_{r,s \notin \lbrace p, q \rbrace} \left\lvert \mel{p q }{}{ r s} \right\rvert & p \neq q \\
0 & p=q
\end{cases}
\end{equation}
Unlike in the original factorization, the vector $\mathbf{S}$ is normalized, as follows:
\begin{equation}
S_p = \frac{\sum_{q} D_{pq}}{\sum_{p,q} D_{pq}}
\end{equation}
We found that this normalization was necessary to eliminate large fluctuations in elements of the matrix $\mathbf{B}$. Additionally, a vector $\mathbf{X}$ of normalization factors for exchange integrals is calculated, with elements defined as:
\begin{equation}
X_i = \sum_a \left\lvert \langle i a | a i \rangle \right\rvert^{1/2}
\end{equation}
The calculation of these normalization factors does not contribute appreciably to the overall computational cost, since they need only be calculated once and stored at the beginning of each simulation.

In our description of the HB-PP scheme in ref \citenum{Greene2019}, the row spaces of the matrix factors of $\mathbf{Q}$ are composed of elements corresponding to single and double excitations from Slater determinants as well as ``no excitation'' elements corresponding to diagonal elements in $\mathbf{P}^{(\tau)}$. These ``no-excitation'' elements are not included in the matrices here, since we altered our description of the factorization scheme to apply only to the off-diagonal part of $\mathbf{P}^{(\tau)}$.

The row space of the matrix $\mathbf{Q}^{(1)}$ consists of generic single and double excitations from each Slater determinant. Elements for single excitations are calculated as
\begin{equation}
Q^{(1)}_{( K, 1 ),J} = \delta_{KJ} \frac{n_\text{s}}{n_\text{s} + n_\text{d}}
\end{equation}
where $n_\text{s}$ and $n_\text{d}$ denote the number of symmetry-allowed single and double excitations, respectively, from the Hartree-Fock determinant. Elements for double excitations are calculated similarly, as
\begin{equation}
\label{eq:q1el}
Q^{(1)}_{( K, 2 ),J} = \delta_{KJ} \frac{n_\text{d}}{n_\text{s} + n_\text{d}}
\end{equation}

Single excitation elements in the remaining matrices in the factorization are defined as in ref \citenum{Greene2019}, so they will not be discussed further here. Elements in $\mathbf{Q}^{(2)}$ for double excitations are specified differently, as
\begin{equation}
Q^{(2)}_{( K, 2, i ), ( K, 2 )} = {S_i}
\end{equation}
where $i$ is constrained to be any of the occupied orbitals in $|K \rangle$ \textit{except the first}. This constraint will be important for enforcing an order among the orbitals involved in double excitations. Notably, the elements for the occupied orbitals in each determinant are not normalized.

Elements in $\mathbf{Q}^{(3)}$ correspond to the second occupied orbital in each double excitation:
\begin{equation}
Q^{(3)}_{( K, 2, i, j ), ( K, 2, i )} = \frac{D_{ij}}{S_i}
\end{equation}
The index of the second occupied orbital in the excitation $(j)$ is restricted to be less than that of the first $(i)$ in order to enforce an ordering between these two orbitals.

Elements in $\mathbf{Q}^{(4)}$, corresponding to the first virtual orbital in an excitation, are specified as
\begin{equation}
Q^{(4)}_{( K, 2, i, j, a ), ( K, 2, i, j )} = \frac{|\braket{i a }{a i}|^{1/2}}{X_i}
\end{equation}
where the index $a$ denotes any virtual orbital in $| K \rangle$ except the first. Recall that Hartree-Fock exchange integrals $\braket{i a }{a i}$ are zero if the spins of orbitals $i$ and $a$ differ. Elements in $\mathbf{Q}^{(5)}$, corresponding to the second virtual orbital $b$, are defined similarly:
\begin{equation}
\label{eq:q5el}
Q^{(5)}_{( K, 2, i, j, a, b ), ( K, 2, i, j, a )} = \frac{|\braket{j b }{ b j}|^{1/2} \delta_{\Gamma_b \otimes \Gamma_a, \Gamma_i \otimes \Gamma_j}}{X_b}
\end{equation}
The orbital $b$ is constrained to be less than $a$ and obey the following symmetry relation:
\begin{equation}
\Gamma_i \otimes \Gamma_j = \Gamma_a \otimes \Gamma_b
\end{equation}
where $\Gamma_x$ denotes the irreducible representation of orbital $x$. This symmetry condition is described in more detail in refs \citenum{Booth2014} and \citenum{Greene2019}.

\section{Parameters for Preliminary Calculations on More Difficult Systems}
\label{sec:hardParams}
This section describes the parameters used to perform the calculations described in Section \ref{sec:concl}. A cc-pVQZ single particle basis was used for the Ne atom. Core electrons were not frozen, so the dimension of the relevant FCI space (10 electrons in 55 spatial orbitals) is $1.51 \times 10^{12}$. The initiator approximation was applied with a threshold of $n_a = 1$. Calculations were executed for 1 million iterations with $m=500,000$ nonzero elements retained in each compression operation. Trajectories were initialized from the Hartree-Fock unit vector.

Calculations on the stretched \ce{N2} molecule were performed at an internuclear distance of 4.2 $a_0$, following ref \citenum{Booth2009}, in a cc-pVDZ basis. The 4 core electrons were frozen, yielding an FCI dimension of 541 million (10 electrons in 26 orbitals). Calculations were executed with $m = 5$ million nonzero elements for 400,000 iterations. Trajectories were initialized from the CISD unit vector. Unlike for the other systems discussed above, the statistical efficiency for this system increased as the initiator threshold was increased beyond $n_a = 1$, up to $n_a = 3$, so a threshold of $n_a = 3$ was used in these calculations. Only the alternative HB-PP factorization was used.

\begin{acknowledgments}
We thank Aaron Dinner, George Booth, and Michael Lindsey for useful conversations and Benjamin Pritchard for his suggestions for improving the performance and readability of our source code. We thank Verena Neufeld for useful comments on the manuscript. S. M. G. was supported by start-up funds from the University of Chicago and by a software fellowship from the Molecular Sciences Software Institute, which is funded by U.S. National Science Foundation grant OAC-1547580. R. J. W. was supported by the National Science Foundation through award DMS-1646339. J. W. was supported by the Advanced Scientific Computing Research Program within the DOE Office of Science through award DE-SC0020427. The Flatiron Institute is a division of the Simons Foundation. Computational resources were provided by the University of Chicago Research Computing Center and the New York University High Performance Computing Center.
\end{acknowledgments}

\section*{References}
\bibliography{Sams_refs,software}

\end{document}